Educational Implications of the 'Self-Made Worldview' Concept

Alexandra Maland and Liane Gabora

Chapter for
*Dynamic Perspectives on Creativity: New Directions for Theory, Research, and Practice in Education*

Published by Springer

Edited by Ronald A. Beghetto and Giovanni E. Corazza

Address for correspondence:
Liane Gabora
Department of Psychology, Fipke Centre for Innovative Research
3247 University Way
Kelowna BC, Canada V1V 1V7
Email: liane.gabora [at] ubc.ca
Tel: 250-807-9849

ABSTRACT

Immersion in a creative task can be an intimate experience. It can feel like a mystery: intangible, inexplicable, and beyond the reach of science. However, science is making exciting headway into understanding creativity. While the mind of a highly uncreative individual consists of a collection of items accumulated through direct experience and enculturation, the mind of a creative individual is self-organizing and self-mending; thus, experiences and items of cultural knowledge are thought through from different perspectives such that they cohere together into a loosely integrated whole. The reweaving of items in memory is elicited by perturbations: experiences that increase psychological entropy because they are inconsistent with one's web of understandings. The process of responding to one perturbation often leads to other perturbations, i.e., other inconsistencies in one's web of understandings. Creative thinking often requires the capacity to shift between divergent and convergent modes of thought in response to the ever-changing demands of the creative task. Since uncreative individuals can reap the benefits of creativity by imitating creators, using their inventions, or purchasing their artworks, it is not necessary that everyone be creative. Agent based computer models of cultural evolution suggest that society functions best with a mixture of creative and uncreative individuals. The ideal ratio of creativity to imitation increases in times of change, such as we are experiencing now. Therefore it is important to educate the next generation in ways that foster creativity. The chapter concludes with suggestions for how educational systems can cultivate creativity.

# CONTENTS

Immersion in a creative task can be an intimate experience. It can feel like a mystery: intangible, inexplicable, and beyond the reach of science. However, science is now making exciting headway into understanding creativity. This chapter examines factors involved in the making of a creative mind, with the ultimate aim of helping educational systems to *cultivate* creative thinking. We believe the time for this is ripe, as our world is changing quickly, yielding new opportunities to be mined, new challenges to be met, and new problems to be solved. Thus, innovative thinkers are needed, perhaps more than ever before.

## The 'Self-Made' Worldview (SMW)

We will use the term *worldview* to refer to a mind as it is experienced subjectively, from the inside. It is a way of *seeing* the world and *being in* the world that emerges as a result of the structure of ones' web of understandings, beliefs, and attitudes. A worldview reveals itself through behaviour, expression, and how it responds to situations (Gabora, 2017a).

The worldviews of uncreative people consist largely of knowledge, social rules, and norms they've picked up from others through imitation and other *social learning* processes (i.e., process that involve learning from others). Uncreative people tend on the whole to be pleasant, reliable, and uncomplicated (Feist, 1998). Their worldviews are largely compilations of the knowledge, rules, and norms they've picked up from the world around them. If you were to envision the worldview of a very uncreative person, it might look like a stack of discrete and separate cards, with each card representing something they've seen, heard, or been told. Though in the classroom this type of student may be a teacher's dream, simply reproducing knowledge while remaining obedient and passive in class may not be what it takes to nurture a future innovator.

The worldview of a creative person, on the other hand, might be envisioned as a stack of cards set on fire, edges ruffling. They are in a state of frenzy, becoming increasingly less separate and distinct; turning into something quite different from what they were previously. Despite the diversity of creative people, they tend to exhibit an overlap of certain personality traits, such as curiosity, openness, high energy, confidence, lack of reliability, rebelliousness, and a tendency to deeply immerse themselves in projects they feel passionate about, often to the exclusion of everything else (Feist, 1998). Unfortunately, teachers often have biases against creativity in the classroom (Beghetto, 2013), as they devalue those types of personality attributes, which may be regarded as disruptive, and are often at odds with the overall structure of modern education.

Of course, in reality, almost everyone falls somewhere in between these two extremes. However, for simplicity, we refer to the worldview of an uncreative person as a *socially-made worldview* and the worldview of a creative person as a *self-made worldview,* or SMW for short.

## The Making of a Self-made Worldview

The above-mentioned process of social learning is sometimes contrasted with *individual learning*, which involves learning outside a social context and which unfolds through direct experience in the world. It is sometimes assumed that individual learning encompasses creativity, but while individual learning culminates in the *discovery* of something in the world, creativity culminates in the *generation* of something that did not previously exist. Moreover, creativity can take place independent of direct experience in the world; it can be wholly or partially internal as opposed to external, and unlike individual learning it can occur in a social context. Thus, creative people venture beyond both social and individual learning; they rehash and reflect upon learned information, put their own spin on it, and make it their own (Feinstein, 2006). This happens through processes that Piaget (2013) referred to as *assimilation*—fitting new information into one's existing web of understandings—and *accommodation*—the complementary process of restructuring one's existing web of understandings to make sense of the new information.

In short, while an uncreative worldview reflects *what the individual has been told*, the structure of a creative worldview reflects *what the individual has done with what he or she has been told:* their imaginative spin, as well as symbol manipulation and deductive and abductive processes. An 'uncreative' student might fit well in a traditional education system wherein fact reproduction and rote memorization make for structured classrooms and efficient evaluations. A 'creative' student, however, is more likely to thrive in an environment that allows for more open-ended teaching methods. (A concrete example of an open-ended assignment will be provided at the end of this chapter.)

A worldview—and particularly a SMW—is *self-organizing* and *self-mending* (Gabora & Merrifield, 2012; see also Osgood & Tannenbaum, 1955). SMWs explore and play with previously encoded knowledge and experiences, looking at them from different perspectives. As such, over time, items in memory tend to weave into a coherent web and take a form that may be unrecognizable to the original. In response to *perturbations*—i.e., external stimuli or internal realizations that are found to be disturbing or inconsistent with deep-seated beliefs—the SMW reconfigures itself through local interactions amongst its components, and thereby maintains a more or less orderly structure. The SMW explores relationships amongst its contents to establish internal consistency, and does so spontaneously. If something surprises us, or someone does something that seems inconsistent with how they normally behave, we cannot stop ourselves from trying to figure out *how* and *why*. Our worldview attempts to mend itself, just as does a body when injured; thus, we are inclined to explore possibilities and modify our interpretations of situations in order to restore our sense of consistency and integrity (Greenwald, Banaji, Rudman, Farnham, Nosek, & Mellott, 2002).

The greater the tendency of a worldview to self-organize, the more likely that it comes to deviate from the status quo. However, since the internal upheaval that comes with self-organized restructuring can be cognitively and emotionally demanding, the SMW individual may be more likely to been seen as unusual (Feist, 1998), or even diagnosed with an affective disorder (Andreasen, 1987). On the plus side though, they have greater potential to produce something—be it an idea, artifact, or class project—that others will regard as creative.

**The Trade-off between Creating and Copying**

It isn't actually necessary for everyone to be creative for the benefits of creativity to be felt by all. We can reap the rewards of a creative person's ideas by copying them, buying from them, or simply admiring them. Few of us can design a skyscraper or compose an opera, but they are nevertheless ours to use and enjoy.

As in any kind of evolutionary process, novelty must be balanced with preservation. In cultural evolution, the novelty-generating component is creativity, and the novelty-preserving components include imitation and other forms of social learning (Gabora, 2013). In an agent-based computational model of cultural evolution (Gabora, 1995), in which artificial neural network-based agents invent and imitate ideas, the society's ideas evolve most quickly when there is a good mix of creative "inventors" and conforming "imitators" (Gabora & Tseng, 2017). Of course, if there are no creative agents, then there are no novel items or ideas to preserve in society. But, if there are too many creative agents, then the collective suffers by never implementing or sharing the fittest of those items or ideas. Thus, a society thrives when some individuals create and others preserve their best outputs.

So, there is a trade-off to peppering the world with creative minds. However, we should not avoid trying to enhance creativity in society—beginning with the classroom—because the pace of cultural change is accelerating more quickly than ever before. In some biological systems, when the environment is changing quickly, the mutation rate goes up. Similarly, in order to generate the innovative ideas that will keep us afloat during times of societal change, we need to encourage individuals to explore their creative musings.

**'Minding the Gap'**

It is often the case that as soon as you clean one part of your house, such as the kitchen countertops, you notice how dirty other parts are, such as the floors and cupboards. The same principle holds true for your internal world; once you restructure one part of your worldview (e.g., from making a new friend who challenges a stereotype you hold), you glimpse the need to revise other parts as well (such as your judgement of others previously labeled with this stereotype). Similarly, it is widely known amongst scientists that answering one question, or solving one problem, often leads to others. So, ironically, though a SMW spends more time 'fixing' itself, it nevertheless still feels more in need of fixing. Whereas the uncreative worldview may not restructure itself until it experiences an *external* perturbation—i.e., a disturbance in the outside world—the SWM continuously reevaluates its own structure by seeking, or even *generating*, perturbations.

Thus, someone with a SMW is particularly inclined to experience the kind of question, problem, sense of incompletion or disconnect, feeling of curiosity, or need for self-expression that Feinstein (2006) notes tends to precede creativity. It may arise suddenly, or slowly over the course of years, and be either inconsequential or of far-reaching importance, and it has been described as a relatively chaotic cognitive state (Guastello, 2002).

The state can be framed in terms of the concept of *entropy*, a term that comes from thermodynamics and information theory, and refers to the amount of uncertainty and disorder in a system. As self-organizing systems, worldviews minimize internal entropy by continually interacting with and adapting to their environments. As open systems, worldviews capture energy (i.e., knowledge or information) from their environment, use it to maintain partially-stable states, and keep their own entropy low by displacing it into the outside world. Hirsh et al. (2012) use the term psychological entropy to refer to anxiety-provoking uncertainty, which they claim humans attempt to keep at a manageable level. However, uncertainty can also be experienced positively, as a wellspring for creativity; not just negatively, as anxiety (Gabora, 2017). (It can also be experienced as some combination of the two.) Accordingly, the concept of psychological entropy has been redefined in terms of *arousal* rather than anxiety, i.e., as arousal-provoking uncertainty rather than anxiety-provoking uncertainty.

Our attention is often pulled toward arenas of life where there is conflict, confusion, or uncertainty, because they are an arousing source of psychological entropy, and the creative process restructures this high entropy material so as to make it more acceptable. Since SMWs are continuously reorganizing and re-assessing themselves, they are more inclined to detect psychological entropy and engage in self-organized restructuring.

**The Chaining of Thoughts and Actions**

Given that a mind is self-organizes, how does it accomplish this? It has been proposed that the extraordinary creativity of the human mind was due to the onset of two exclusively human cognitive abilities. The first, which evolved following an increase in cranial capacity (skull size) approximately two million years ago, was the onset of the *self-triggered recall and rehearsal* loop (Donald, 1991). This 'loop' marked the capacity for one thought to trigger another thought and enabled our early ancestors to rehearse and refine skills, recall and reflect upon the past, and fantasize about the future. It has also been referred to as *chaining* (Gabora & Tseng, 2017), as it enables the chaining of thoughts and actions into streams of free-association, critical reflection, or complex behavioral sequences. Chaining enabled our ancestors to minimize arousal by repeatedly considering high psychological entropy material from new contexts until it was amply reorganized and a new idea or perspective was achieved.

Chaining concepts began by connecting items in memory that were previously assumed to be unrelated. This enabled our ancestors to respond flexibly to stimuli and then reflect on their responses. The hypothesis that the onset of the capacity for chaining made cultural evolution possible can be demonstrated by an agent-based computer model of cultural evolution, called EVOC (for EVOlution of Culture). In EVOC, chaining increases the mean fitness and diversity of cultural outputs, which supports the hypothesis that chaining transformed a culturally stagnant society into one characterized by open-ended novelty (Gabora, Chia, & Firouzi, 2013; Gabora & Smith, submitted).

Although the ability to chain began long ago, it is probably not the case that all modern people are equally inclined to engage in it. The greater the extent to which one chains thoughts and ideas together, the more likely one is to end up with a worldview that veers from the status quo. Conversely, those with a SMW are more likely to engage in chaining, weaving stories about the past or fantasizing about the future. This is consistent with the notion of an absent-minded professor or a daydreaming student; such an individual tends to be 'lost in thought' as opposed to living in the present moment. However, if something *does* manage to pull their attention, it tends to be more thoroughly honed before they settle into a particular understanding of it.

**Shifting Between Analytic and Associative Modes of Thought: Contextual Focus**
The second uniquely human cognitive ability is posited to have given rise to the 'big bang' of human creativity (Mithen, 1998) approximately 60,000 years ago. It is referred to as *contextual focus* (CF) because it involves the capacity to adapt ones' mode of thought to the context by focusing or defocusing attention (Gabora, 2003; Chrusch & Gabora, 2014). CF enabled our ancestors to spontaneously shift between a *convergent* or *analytic* mode of thought and a *divergent* or *associative* mode of thought, wherein the interconnections between concepts and ideas become more fluid and malleable. Convergent or analytic thought is conducive to mentally demanding analytic tasks; it stifles associations, and reserves mental effort for symbol manipulation and detecting relationships of *causation*. Divergent or associative thought, on the other hand, is conducive to mind-wandering, forging new associations when stuck in a rut, and detecting relationships of *correlation*.

While chaining enabled the connecting of *closely* related items in memory, CF enabled the forging of *distant* connections. Divergent thought allowed these connections to be glimpsed, and convergent thought enabled them to be polished into their final form.

It is interesting to view these forms of thought not just in terms of what they're good for but in terms of how they process information at the conceptual level, i.e., the level of concepts. While convergent thought involves using concepts in their most stereotypical, undifferentiated form, i.e., *sticking to their most conventional contexts*, divergent thought is characterized by exploring the broader 'halo' of *potentiality* surrounding concepts, *i.e., the new meanings or feelings that arise when concepts are conceived of in particular, often unconventional, contexts* (Gabora, 2018). Divergent thought is a matter of making broad use of contextual information to make associations in *constrained* (though potentially obscure) ways, rather than *loosely* expanding ones' sphere of associations in a generic sense. Thus, while in a convergent mode of thought you might think of a toothbrush only in terms of its role in brushing teeth, but while puzzling over how to clean a narrow bottle, in a divergent mode of thought you might conceive of a toothbrush as a possible bottle cleaner.

CF, like chaining, contributes to the making of a SMW, and does so through the forging of strange and often personal associations. The formation of a SMW entails the ability to not just *use* both modes of thought, but apply them when needed. As the SMW individual experiences more psychological entropy, there is a greater perceived need for restructuring, and thus a need for CF. When psychological entropy is experienced, creative individuals attempt to mend their fractured worldview by making use of divergent thought, which can forge new connections and come up with ideas. However, since new ideas are born in a divergent mode of thought, they are not likely to be immediately implementable, or even intelligible. Novel ideas and discoveries must be reflected upon or honed from different perspectives in a more clear-headed, convergent mode of thought. In short, the SMW is able to transfer what is discovered in one mode of thought to the other, thereby achieving a more nuanced web of associations.

Like chaining, CF has also been modeled using EVOC (Gabora, Chia, & Firouzi, 2013; Gabora & Smith, submitted). The mean fitness of actions across EVOC's artificial society increased with CF, and CF was also particularly effective when the fitness function changed. These findings support CF's hypothesized usefulness in breaking out of a rut, adapting to new or changing environments, and generating insight.

**From Task or Problem, to 'Half-Baked' Idea, to Creative Output**
Honing is the process of putting chaining and CF to work in the generation of a creative output. Honing involves viewing something in a new context, which leads to a new take on it, which suggests another new context to consider it from, and so forth, until psychological entropy reaches an adequately low level, the gap is filled, and the worldview is less fragmented (Gabora, 2017a). A more stable image of the world, and one's own relation to it, comes into focus through immersion in a creative task (Pelaprat & Cole, 2011). Honing may also utilize *imagination:* the ability to form internal images or ideas by combining previous knowledge and experience.

A SMW is likely to experience cognitive states that feel unclear, ill-defined, and in need of honing, because it is in a process of constant renewal. When a problem (or task or concept) is considered in a new way for the first time, it is not always clear how the problem and context fit together. The phrase *half-baked idea* epitomizes how a creator may not even be able to comprehend their own newly hatched idea. Even so, since the SMW is more likely to *have* half-baked ideas in the first place, it is more likely to *hone* them into a state in which they are able to manifest, or take on a life in the outside world. Thus, the more half-baked ideas, the more likely one of them will evolve into something new and useful.

Our minds gain a deeper understanding of something, such as a half-baked idea, by looking at it from different perspectives. It is this kind of deep processing and the resulting integrated webs of understanding that make the crucial connections that lead to important advances and innovations (Gabora, 2017a). Encouraging individuals to engage in deeper reflection and understanding is vital to creative thinking, and is particularly important in our society's high-stimulation environment. Youth spend so much time processing new stimuli that there is less time to venture deep into stimuli they've already encountered. With smartphones in hand and a continuous inflow of new content to consume, there is less time for thinking about ideas and situations from different perspectives such that ideas are honed and worldviews become more integrated.

**Insight as Self-Organized Criticality**

*Insight* can be experienced as an '*aha!*' moment and defined as a sudden new representation of a task or a realization of how to go about it (Mayer, 1995). It is a deliberate cognitive reorganization that concludes in a new and useful interpretation or understanding that is non-obvious, or even peculiar (Kounios & Beeman, 2009, 2014).

Insight marks a cognitive phase transition (Stephen, Boncoddo, Magnuson, & Dixon, 2009) arising due to *self-organized criticality*, or SOC (Gabora, 1998). SOC is a phenomenon wherein, through simple local interactions, complex systems tend to find a critical state—at the cusp of a transition between order and chaos—from which a single minor agitation sometimes exerts a disproportionately large effect (Bak, Tang, & Weisenfeld, 1988). The signature of SOC is an inverse power law relationship in which most perturbations have little effect, but the occasional perturbation has a dramatic effect. For example, most of our moment-to-moment thoughts are inconsequential, but once in a while one thought triggers another, which triggers another, causing an 'avalanche' of reconceptualizations, concluding in a noticeably altered understanding of a concept, i.e., a moment of insight.

Like other SOC systems, a creative mind may function within a system midway between order (involving the systematic progression of thoughts), and chaos (wherein everything reminds you of everything else) (Gabora, 1998). The SMW may make use of CF to *stay* in this regime, by utilizing systematic analysis when life gets complicated, and letting the mind wander when things get dull. As such, when our learning environments are optimally designed to allow CF to be utilized, SMWs and the creative minds that house them have room to exercise.

Some attribute the unpredictability of insight to a blind, sequential, trial-and-error process of idea generation, but insight actually arises through SOC due to the dynamical process of constant change and interaction in SOC systems. This claim is consistent with various lines of evidence across various domains of inquiry. Firstly, insight as SOC corroborates with findings that a series of small conceptual changes is often followed by large-scale creative conceptual change (Ward, Smith, & Vaid, 1997), and with the fact that worldviews rapidly reconfigure in response to external inputs. It is also consistent with findings that insight tends to come suddenly (Gick & Lockhart, 1995) and with ease (Topolinski & Reber, 2010), despite often succeeding intensive effort. Evidence that power laws are applicable to novelty transmission is another indication of SOC (Jacobsen & Guastello, 2011). Further, there is evidence of SOC in the human brain (Kitzbichler, Smith, Christensen, & Bullmore, 2009), and at the cognitive level. For example, word association studies have shown that concepts are clustered and sparsely connected, with some having many associates and others few (Nelson, McEvoy, & Schreiber, 2004). Semantic networks exhibit the sparse connectivity, short average path lengths, and strong local clustering that are typical of SOC structures (Steyvers & Tenenbaum, 2005).

### A Micro-Level View of the Self-made Worldview

Having introduced the concept of a SMW as it manifests at the macro-level of everyday life, let us now briefly examine how a SMW is structured at the micro-level of neuron assemblies.

**The Birthplace of Creativity**

The notion of a SMW can be grounded in the brain's architecture by examining some key attributes of associative memory. First, our memory encoding is *sparse* (Kanerva, 1988). This means that the number of possible items that can be encoded is far greater than the number of neurons in the human brain.

Second, human memories exhibit *coarse coding* (Sutton, 1996): they are encoded in neurons that are sensitive to ranges of microfeatures. When memory organization follows coarse coding, a neuron responds maximally to a particular microfeature, and responds to a lesser extent to similar microfeatures. For example, neuron *A* may respond preferentially to a certain color (e.g., pink), while its neighbor *B* responds preferentially to a slightly different color (e.g., magenta), and so forth. However, although *A* may respond maximally to pink, it still responds, to a lesser degree, to magenta (or another similar color).

The upshot of coarse coding is that an item in memory is *distributed* across a cell assembly of neurons that are sensitive to particular properties. This means that a given experience activates not just one neuron, nor every neuron to an equal degree, but spreads across an assembly. Thus, the content of an item in memory doesn't arise due to the activation of any single neuron, but an overall pattern of activation, wherein the same neurons are used and re-used in different capacities (McClelland, Rumelhart, & Hinton, 2003; Edelman, 1987). Distribution guarantees the possibility of forging associations amongst items (Josselyn, et al., 2015)—whether closely or distantly related—in a potentially new and useful, even surprising, way.

Human memory is also *content addressable*: there is a systematic relationship between stimulus content and the cell assemblies that encode it (Kanerva, 1988). This enables the brain to take what is being experienced (i.e., context), and then naturally bring items to mind that are related, in either obvious or unexpected ways. Content addressability ensures that memory items are awakened by stimuli that are similar, or that resonate with one another (Hebb, 1949). It is often the case that in moments of insight, the correlation between items has never been explicitly noticed, and their overlapping distribution is what awakened this new connection.

It is because of these features of memory that we are able to essentially create a new concept, or draw a new connection, that is more appropriate to the situation than anything our brain has ever been fed as input (McClelland, Rumelhart, & Hinton, 2003). If memory were not sparse, and all neurons were packed like sardines within a small region, then we would not be able to make distinctions across the vast range of stimulus qualities that we can. Without distribution and coarse coding, there would be no overlap between items that share microfeatures, and thus no means of forging associations between them. Without content addressability, these associations would not be appropriate or significant; we would be wandering blindfolded through memory (McClelland, Rumelhart, & Hinton, 2003). This cognitive architecture, as well as the compulsion of a SMW to self-organize in response to psychological entropy, is the key to our creativity.

Lastly, stored items in memory are never re-experienced in exactly the same form as when first encoded, because the meanings of items are not stored in seclusion, and are in part derived from the meanings of other representations that excite associated groups of neurons. Thus, retrieving a specific, singular memory is essentially impossible and memory does not work through a verbatim retrieval process. Recalled items are inevitably shaped by experiences the individual has had since encoding (Paterson, Kemp, & Forgas, 2009; Schacter, 2001), because the strength and pattern of connections amongst neurons are always changing (McClelland, 2011). Changing patterns also means that items can spontaneously re-assemble in a way that relates to the task at hand. In this way, a worldview gets restructured.

**How Chaining is Implemented at the Neural Level**

At the neural level, the onset of chaining meant that more features of any particular experience were encoded in memory, because more microfeature-specific neurons participated. It follows that the proclivity for a SMW may stem from the tendency to encode items in greater detail. Since more details lead to more distinctions, which enable more routes by which one thought can evoke another, this ultimately leads to more possibilities for connecting the dots between past and present experiences. Altogether, chaining and detailed encoding produce a more nuanced worldview. This is why our modern, highly stimulating world, in which classrooms expect efficient, rehearsed answers, may be at a disservice to creative minds—which are born out of attention to detail and thorough investigation.

## How Contextual Focus is Implemented at the Neural Level

Recall, CF is the capacity to effectively tailor one's mode of the thought to the current situation or context by focusing or defocusing attention (Gabora, 2003). Analytic thought involves focusing attention and constraining activation such that items are considered in a compact form that is amenable to complex mental operations. Associative thought involves defocusing attention such that obscure, though potentially relevant, aspects of a situation to come into play, and the set of possible associations expands.

It has been proposed that the 'shift' experienced in CF is carried out through the recruitment and decruitment of *neurds* (Gabora, 2010, 2018; Gabora & Ranjan, 2013): the neural assemblies that are *not* activated in analytic thought, but *are* in associative thought (Gabora, 2010; see also Ellamil, Dobson, Beeman, & Christoff, 2012; Yoruk & Runco, 2014). Neurds respond to properties that are of minimal relevance to the current thought, and thus more distantly associated. The subset of cell assemblies that count as neurds shifts depending on the situation, and so they do not reside in any particular region of memory. Every different perspective one takes toward a particular thought, idea, or concept recruits a different set of neurds.

In associative thought, diffuse activation recruits more cell assemblies, including neurds, which enables one thought to stray far from another, while still retaining a thread continuity. Insight occurs when concepts previously assumed to be unrelated are united and sufficiently activated to cross into conscious awareness (Salvi, Bricolo, Kounios, Bowden, & Beeman, 2016). Following insight, an idea must be honed to completion, which necessitates the shift to an analytic mode of thought. This analytic shift is accomplished by dismissing the cell assemblies that are active during associative thought, i.e., the *decruitment* of neurds.

An individual is more likely to have a SMW if they utilize CF to tailor their mode of thought to the current situation to a greater extent. Although engagement in associative thought navigates memory both deeply and quickly, the ability to reign in it in is as important as the ability to initiate it. If everything always reminds you of everything else all the time, then the ability to draw connections becomes counterproductive. The ability to shift between, and control, both associative and analytic thought enables the structure of one's worldview to capture not only the contents acquired through social and individual learning, but also how they are all related. These relationships include not only basic-level connections, but relationships with respect to different contexts, and at different levels of abstraction.

## The Ancestral Origins of New Ideas

The fact that associations come to mind spontaneously as a result of representational overlap and shared features means there is no need for memory to be randomly searched to make creative associations (Gabora, 2010). The more detail with which stimuli and experiences are encoded, the greater their distributed representations overlap, and the more potential routes by which they can act as contexts for one another and combine.

Let us consider a concrete example. One morning, one of the authors of this chapter awoke from a dream in which she opened a small business basically hauling people around in a bus, and had come up with the slogan, "We put the Bus in Business". Where on earth did *that* come from, she wondered. It was obvious, though. It came from a combination of the Sense8 Netflix series she'd been binge-watching in which one of the heros hauls people around in a bus, the dystopian science fiction she'd been reading, and sayings like "We put the Fun in Fundraiser".

You may now have some idea of what caused her mind to pull these things together to generate this weird, new slogan. In the context of wanting to invent a slogan about a 'bus business' her dreaming brain didn't randomly mutate BUSINESS to get, say, "we put the harness in business"; it immediately hit on something that, albeit silly, is at least relevant (even when dreaming!). This is consistent with the distributed and content-addressable architecture of memory; as a result, associations are forged by way of shared structure. The first author's dreaming mind connected BUS and BUSINESS on the basis of the shared syllable BUS, because they both activate neurds that are tuned to respond to the features of this syllable.

Having examined how new ideas emerge in associative memory, it is clear that they do not 'arise out of the blue'; they bear some relationship to knowledge and experiences encoded in memory before the creative act took place. Thus, due to the organization of memory, insight is not simply a matter of chance or expertise, but of becoming newly aware of the shared features between memories and concepts.

## The Evolution of Creative Worldviews

To understand the mystery that is creativity, we must assess where creativity fits into the 'big picture' of humanity. Humans uniquely participate in two evolutionary processes: biological and cultural. Within this second cultural evolutionary process creativity is the driving force that enables cultural outputs to proliferate without limits (Gabora, 1997).

Although it may at first seem possible that culture, like biological organisms, could evolve through a Darwinian process (of natural selection or 'survival of the fittest'), there is abundant theoretical and empirical evidence against a Darwinian framework for culture (Gabora, 2004, 2006, 2011, 2013). Even very early life itself went through a stage where it evolved—i.e., exhibited cumulative, adaptive, open-ended change—through a process that functioned rather differently from natural selection. What made early *biological* evolution possible was the emergence of *protocells*: simple cell-like structures that were self-organizing, self-mending, communally interacting, and self-reproducing (Hordijk, Hein, & Steel, 2010; Kauffman, 1993). It has been shown that these protocells evolved through a more haphazard process referred to as *communal exchange*, which involved interactions not just within but amongst these self-organizing structures (Vetsigian, Woese, & Goldenfeld, 2006).

It has been proposed that what made *cultural* evolution possible is the emergence of *mind or worldview* that is self-organizing, self-mending, communally interacting, and self-reproducing (Gabora, 2013). In short, the protocell is to biological evolution what the worldview is to cultural evolution. The internal state of the worldview changes dynamically due to a communal exchange of information with the external world, just like the earliest forms of life. By sharing experiences, ideas, and attitudes with others, individuals can influence the process by which another worldview may form and transform. One of the earliest environments that expose us to sharing through communal exchange is the classroom, where there is endless potential for unique interactions to take place. But if teaching methods follow strict, traditional guidelines in this early environment, then peer interaction may be discouraged, or at least overlooked, during the learning process, thus stifling the sharing and questioning of ideas that construct SMWs.

When people directly or indirectly interact with and make impressions on one another, their worldview's configuration can be passed on by influencing and being imperfectly reconstituted others. For example, students may expose fragments of what was originally their parent's or teacher's worldview according to the student's own unique experiences and physical limitations, thereby forging unique internal models of the relationship between self and world. Thus, worldviews are not just communally interacting, but also self-regenerating. However, there is no universal 'self-assembly code' (e.g., DNA) that ensures the reliability of replication (as in biology). Consequently, particular beliefs or ideas may take on a different role within various worldviews as a result of the experiences and structures unique to that particular individual.

A worldview is more likely to unfold into a SMW if it is exposed to more social and environmental interactions and experiences, thus encoding more information and having more potential configurations. Therefore, a worldview evolves through communal exchange by interweaving both (1) internal interactions amongst its parts, and (2) external interactions with others; a SMW does this to a greater extent. When creativity is viewed in light of its role in fueling cultural evolution, the proposal of insight as self-organized criticality (SOC) fits into a broader conceptual framework. According to the *theory of punctuated equilibrium*, for which there is substantial well-documented evidence, changes in biological species are restricted to rare, rapid events interspersed amongst prolonged periods of stasis (Eldridge & Gould, 1972). Punctuated equilibrium is perhaps the best-known example of SOC in nature, and it has been suggested that cultural evolution, like biological evolution, exhibits punctuated equilibrium (Orsucci, 2008). Insight may play the role in cultural evolution that punctuated equilibrium plays in biological evolution, fueling the reorganization of, not species in an ecosystem, but concepts and ideas in an 'ecology of mind'.

One of the more prominent implications of applying communal exchange to cultural evolution is that it is not the creative artifacts or ideas that are evolving, but the worldviews that produce them (Gabora, 2004). Creative outputs are, in as sense, the byproducts of evolving worldviews. Since a worldview cannot be seen, it reveals itself through behavioural consistencies in social interaction and through generating creative outputs (Gabora, 2017a). For example, when students debate about politics, or present their art projects to each other, these interactions and outputs are what allow worldviews to view and shape one another.

## The Role of Education

Just as the cumulative creativity of biological evolution completely transformed our planet, the cumulative creativity of cultural evolution is completely transforming our planet. Although the creative processes underlying cultural evolution may be more strategic—i.e., less random—than biological evolution, we still cannot predict where its headed. While we have made considerable headway in understanding how the creative process works, we are no better able to predict the next new gadget that will captivate us nor the next novel that will move us to tears. Thus, we cannot predict the world that today's students will have to navigate once they graduate.

We *can*, however, educate them to be, not compilations of knowledge and social rules, but SMWs, able to weave what they are told into a unique cognitive structure of their own making that reflects their unique proclivities and experiences. This means that providing opportunities not just to learn but to *explore* how what they are learning could relate to personal experiences and knowledge in other domains. In this time of not just unprecedented cultural change but change to the planet that sustains us, it is crucial that we educate students to be able to creatively respond to perturbations and cope with the unexpected.

Although teachers claim to value creativity, many hold negative attitudes toward the attributes associated with it, as many fear that encouraging creativity in the classroom could lead to chaos (Beghetto, 2007). Teachers tend to emphasize correct responses and the regurgitation of knowledge at the expense of creativity, as such a framework allows for more straightforward evaluation. Creative projects and testing, on the other hand, take considerable effort and result in minimal benefits to standardized testing. Then, when efforts are made to encourage creativity within an educational setting, neither teacher nor students often knows what the expectations are. However, despite the risks of encouraging creative classrooms, such an investment in education is critically important and potentially rewarding (Ranjan & Gabora, 2012). What is essential to the challenges of our future is of course innovative products and ideas, but also a world where the type of people inhabiting it have benefitted from the creative process.

This notion of potentiality applies not just to concepts but also to students. Each student can be seen as a unique wellspring of creative potentiality, which the teacher can help actualize, or bring to fruition. Just as exploring the 'halo' of potentiality surrounding ideas and concepts can lead to new, unexpected perspectives and associations, the thoughts and ideas of a student—as he or she interacts with a teacher, particular lesson, or approach to teaching it—can potentially follow new, creative trajectories. Allowing the full-ranging exploration of a student's potentiality affords the possibility of new emergent outcomes that could not be predicted in a straightforward logical way from knowledge of the teacher, student, or lesson plan (Ranjan & Gabora, 2012).

## Cultivating Creativity in the Classroom

An obvious final question to address is: how can creativity be nurtured in an educational setting? Incorporating creativity into the classroom may appear challenging and counterproductive, given that the current system is rooted in a certain structure, but there are a multitude of ways to do so (see Gregerson, Kaufman & Snyder, 2013). The following section outlines three key ways in which teachers can begin to incorporate creativity (Gabora, 2017b).

First, it is essential to focus less on the reproduction of information and more on critical thinking and problem solving. Doing so allows students to build the necessary skills to engage in CF, wherein associative thought processes allow them to traverse their minds for ideas, and then analytic thought refines what is found. As discussed above, CF is an important brick in the foundation of a SMW. Incorporating this is as simple as using open-ended written questions as opposed to multiple choice questions on exams, or at least having a mixture. For example, presenting a real-world scenario and then following up with a question that requires using knowledge learned in class, as well as critical thinking and reflective skills.

Second, teachers can curate activities that transcend traditional disciplinary boundaries, such as by painting murals that depict biological food chains, or acting out plays about historical events, or writing poems about the cosmos. After all, the world doesn't come carved up into different subject areas. If our culture and classrooms tell students that these disciplinary boundaries are real, then their thinking becomes trapped in them. In reality, harsh lines do not exist between various domains in life, as the products of any creative act (be it writing a song or solving a math problem) are not the core elements of cultural evolution, but the ever-changing worldviews are. A worldview that is dynamic and creative may be reflected in a student who is capable of both building a beautiful diorama of a cell body and correctly labelling all of the parts.

Third, by posing questions and challenges, and then following up with opportunities for solitude and reflection, teachers can provide time and space for more detailed contemplation. In the current high-stimulation environment that students are exposed to daily, classrooms have the potential to offer a place for deeper reflection and understanding. Reflecting upon what is learned or asked fosters the forging of new connections that is so vital to creativity. While time-constraints may lead students to produce uncreative answers that reflect what they've been told, allowing time for information to 'incubate' may cultivate more creative outputs that showcase what students have done with what they've been told. It may be especially helpful for teachers or professors to view their students as entities that are able to rehash and reflect upon learned information, put their own spin on it, and later, do something novel with it (Ranjan & Gabora, 2012). These types of people are more likely to end up with a creative SMW.

## A Creative Assignment in Practice

To make the above suggestions more concrete, let us walk you through an example of a university classroom assignment (from Ranjan & Gabora, 2012)—aspects of which could be made to suit a younger group, as well. At the beginning of the term, students are asked to choose to complete a presentation, essay, or project, which is due at the end of the term. Students may choose their own topic, but it must pertain to a major theme or subject of the course. By giving students a choice of format and topic, plus the entire term to work on it, they can discover and explore their own interests and talents in-depth, and mull over the assignment from day one. Although it may be difficult to initially generate a topic, the time allowance means that a student can engage in associative thought and connect their own curiosity and abilities with what has been learned in class. In this associate stage of idea-generation, one thought is able to stray far from the next, yet a thread continuity is still retained by the subject or topic (e.g., politics or biology), keeping even far-reaching ideas somewhat relevant to the assignment. Once some half-baked ideas are conjured up, analytic thought constrains the student's flight of ideas so that they can choose which one best fits the course content and will produce a learning experience tailored to the students' interests. Here, students are learning to engage in CF and use it to their advantage.

Since creativity thrives in situations where both freedom and constraint coexist, for an open-ended task such as this, instructors can ensure that students have clear expectations by both providing examples and breaking up the assignment into multiple steps. For example, they could be asked to submit just the title, one sentence, and one reference two weeks after the start of class, then submit one paragraph one month after the start of class, (and so forth), with feedback provided at each stage. In this way, students have lots of opportunity to hone and revise their ideas, and fulfill their creative potential.

This assignment incorporates (1) focusing on critical thinking and problem solving, (2) posing questions and challenges followed by opportunities for solitude, reflection, and detailed contemplation. It could also potentially incorporate (3) activities that transcend traditional disciplinary boundaries; for example, students might respond to this kind of open-ended assignment by creating a video game that explores the laws of physics, or by writing a short story about someone with a psychological disorder.

Studies on conceptual combinations suggest that because more ambiguous tasks such as this offer more opportunities to for students to connect dissimilar concepts, they may result in more creative outputs (Wilkenfeld & Ward, 2001). When students choose their own topics, they are more likely to be intrinsically motivated, and to find personal meaning in what they are learning. Giving them several months to carry out the assignment, and staggering it into sub-goals, encourages students to put more of themselves into their work, which, and provides sufficient time for half-baked ideas to crystallize into final creative products that students can feel proud of. Although some skills may be specific to particular domains (e.g., multiplication ability is more specific to math than literature), creative processes of the kind that may be engendered by such open-ended assignments may catalyze internal transformation that transcends such boundaries.

## Conclusion

Although educators claim to value creativity, a disconnect exists between this claim and what actually happens in school. This is likely because traditional misconceptions and assumptions about creativity keep educators from effectively incorporating it in their syllabi. Thus, the aim of this chapter was to thoroughly introduce a new conception of the creative mind as a complex, dynamical system that begins with worldviews. The ideas proposed here paint creativity as not being domain-specific, and as not measurable by tangible creative outputs. Creativity can be explored as much in chemistry as it is in graphic design, as the real transformation that takes place is internal. Understanding what a creative worldview looks like and how it is different, is the necessary first step in exploring practical applications. We believe that no other investment in education could be more important and potentially rewarding than encouraging this understanding and thus encouraging creative classrooms. By nurturing the skills and attributes characteristic of creative minds, education systems have the potential to cultivate the kinds of people that society needs.

## Acknowledgments

This work was supported by a grant (62R06523) from the Natural Sciences and Engineering Research Council of Canada.